\documentclass[english]{llncs}
\usepackage[T1]{fontenc}
\usepackage[latin9]{inputenc}
\usepackage{babel}

\usepackage[overlay,absolute]{textpos}

\begin{document}

\begin{textblock*}{10in}(38mm, 10mm)
{\textbf{Ref:} \emph{International Conference on Artificial Neural Networks (ICANN)}, Springer LNCS,}
\end{textblock*}
\begin{textblock*}{10in}(38mm, 15mm)
{Vol.~10614, pp.~91--99, Alghero, Italy, September, 2017.}
\end{textblock*}

\title{DeepAPT: Nation-State APT Attribution Using End-to-End Deep Neural Networks}

\titlerunning{DeepAPT: Nation-State APT Attribution Using End-to-End Deep Neural Networks}

\author{Ishai Rosenberg, Guillaume Sicard, and Eli (Omid) David}

\authorrunning{I. Rosenberg, G. Sicard, and E.O. David}

\institute{
Deep Instinct Ltd \\
\email{ \{ishair,\ guillaumes,\ david\}@deepinstinct.com}
}

\maketitle
\begin{abstract}
In recent years numerous advanced malware, aka \emph{advanced persistent threats} (APT) are allegedly developed by nation-states. The task of attributing an APT to a specific nation-state is extremely challenging for several reasons. Each nation-state has usually more than a single cyber unit that develops such advanced malware, rendering traditional authorship attribution algorithms useless. Furthermore, those APTs use state-of-the-art evasion techniques, making feature extraction challenging. Finally, the dataset of such available APTs is extremely small. 

In this paper we describe how deep neural networks (DNN) could be successfully employed for nation-state APT attribution. We use sandbox reports (recording the behavior of the APT when run dynamically) as raw input for the neural network, allowing the DNN to learn high level feature abstractions of the APTs itself. Using a test set of 1,000 Chinese and Russian developed APTs, we achieved an accuracy rate of 94.6\%. 
\end{abstract}

\section{Introduction}

While malware detection is always a challenging research
topic, a special challenge involves nation-state advanced persistent threats (APT), highly sophisticated and evasive malware. Since the usage of such cyber weapons might be considered an act of war \cite{Hathaway12}, the question ``which country is responsible?'' could become critical.

In this paper we use raw features of dynamic analysis to train
a nation-state APT attribution classifier. The main contribution of this paper is providing the first nation-state APT attribution classifier, which achieves a high accuracy on the largest test set of available nation-state
developed APTs ever collected, successfully attributing new malware families.

The rest of the article is structured as follows: Section 2 contains
the relevant related work to our use cases. Section 3 specifies the
problem definition and the unique challenges in this domain, both
with nation-state APT attribution in general and especially when using
feature engineering. Section 4 contains our nation-state APT attribution
classifier implementation and the experimental results. Section 5
contains our concluding remarks.

\section{Background and Related Work}

There are numerous topics related to authorship attribution, such as plagiarism detection, books authorship attribution, source code authorship attribution and binary program authorship attribution. \cite{Stamatatos09} provides a broad review of many of those topics, including the natural language processing (NLP) and traditional machine learning (ML) algorithms and the relevant features used, for example lexical (e.g., word frequencies), syntactic (e.g., sentence structure), semantic (e.g., synonyms), and application specific (such as a specific structure, for instance HTML). 

In the following sub-sections, we focus only on the ones relevant to our work (and ignoring those that are irrelevant, such as source code authorship attribution, which cannot be used in our case since a source code is not available for the APTs).

\subsection{Binary Code Authorship Attribution}

Certain stylistic features can survive the compilation process and
remain intact in binary code, which leads to the feasibility of authorship
attribution for binary code. Rosenblum et al. \cite{Rosenblum11}
extracted syntax-based and semantic-based features using predefined
templates, such as idioms (sequences of three consecutive instructions),
n-grams, and graphlets. Machine learning techniques are then applied
to rank these features based on their relative correlations with authorship.
Alrabaee et al. \cite{Alrabaee14} extracted a sequence of instructions
with specific semantics and to construct a graph-based on register
manipulation, where a machine learning algorithm is applied afterwards.
Caliskan et al. \cite{Caliskan15} extracted syntactical features present
in source code from decompiled executable binary. 

Though these approaches
represent a great effort in authorship attribution, it should be noted
that they were not applied to real malware. Furthermore, some limitations
could be observed including weak accuracy in the case of multiple
authors, being potentially thwarted by light obfuscation, and their
inability to decouple features related to functionality from those
related to authors styles. 

\subsection{Malware Attribution}

The difficulty in obtaining ground truth labels for samples has led
much work in this area to focus on clustering malware,
and the wide range of obfuscation techniques in common use have led
many researchers to focus on dynamic analysis rather than static features (i.e., instead of examining the static file, focus on the report generated after running the file dynamically in a sandbox).

The work of Pfeffer et al. \cite{Pfeffer12} examines information
obtained via both static and dynamic analysis of malware samples, in order to
organize code samples into lineages that indicate the order in which
samples are derived from each other. Alrabaee et al. \cite{Alrabaee17}
have used both features extracted from the disassembled malware code
(such as idioms) and from the executable itself, used mutual information
and information gain to rank them, and built an SVM classifier using
the top ranked features. Those methods require a large amount of pre-processing
and manual domain-specific feature engineering to obtain the relevant features.

The malware attribution papers mentioned so far are applicable only to cases where a malware is an evolution of another malware (e.g., by mutation), or from
the same family, functionality-wise. These methods are not effective when completely different families of malware are examined. Our paper presents a novel application of DNN for APT attribution, specifying which nation has developed a specific APT, when the APTs in question are not derivatives of one another, and belong to completely different families.

The work of Marquis-Boire et al. \cite{Boire15} examines several
static features intended to provide credible links between executable
malware binary produced by the same authors. However, many of these
features are specific to malware, such as command and control infrastructure
and data exfiltration methods, and the authors note that these features must
be extracted manually. To the best of our knowledge, this is the only
available paper that explicitly dealt with nation-state APTs detection
(using features common in them, such as APTs). However, those use
cases are limited, and no accuracy or other performance measures were provided. In addition, the paper did not deal with classifying which nation developed
the malware, and rather mentioned that one could use the similarities
between a known (labeled) nation-state APT to an unknown one to infer
the attribution of the latter.

\section{Problem Definition: Nation-State APT Attribution}

Given an APT as an executable file, we would like to determine which nation
state developed it. This is a multi-class classification problem, i.e., one label per candidate nation-state.

\subsection{The Challenges of Nation-State Attribution}

Trying to classify the nation that developed an APT can be an extremely
challenging task for several reasons that we cover here.

Each nation-state has usually more than a single cyber unit developing
such products, and there is more than a single developer in each unit.
This means that the accuracy of traditional authorship attribution
algorithms, which associates the author of source code or program
using stylistic features in the source code, or such features that
have survived the compilation, would be very limited.

These APTs also use state-of-the-art evasion techniques, such-as anti-VM,
anti-debugging, code obfuscation and encryption (\cite{Virvilis13}),
making feature extraction challenging.

Moreover, the number of such available APTs is small, since such APTs
tend to be targeted, used for specific purposes (and, unlike common criminal
malware, not for monetary gain) and therefore are not available on
many computers. Their evasion mechanisms make them hard to detect as well.
This results in a further decrease in the training set size from which
to learn.

Finally, since nation states are aware that their APTs can be caught,
they commonly might try to fool the security researchers examining the APT
to think that another malware developer group (e.g., another nation) has developed it (e.g., by adding the APT strings in a foreign language, embedding data associated with a previously published malware, etc.). That is, unlike traditional authorship attribution problems, in this case the ``authors'' are actively trying to evade attribution and encourage false attribution.

Despite these issues, manual nation-state APT attribution is performed,
mostly based on functional similarities, shared command and control servers (C\&Cs, which provide an accurate attribution), etc. For example,
the APTs Duqu, Flame and Gauss were attributed to the same origin
as Stuxnet following a very cumbersome advanced manual analysis (\cite{Bencsath12}). The question is: How can we overcome
these challenges and create an automated classifier (that does not require lengthy manual analysis)?

\subsection{Using Raw Features in DNN Classifications in the Cyber Security Domain}

One of Deep Neural Networks (DNN) greatest advantages is the ability
to use raw features as input, while learning higher level features
on its own during the training process. In this process, each hidden
layer extracts higher level features from the previous layer, creating
a hierarchy of higher-level features.

This is the reason why deep learning classifiers perform better than traditional machine learning classifiers in complex tasks that requires domain-specific
features such as language understanding \cite{Collobert11}, speech recognition, image recognition
\cite{Zeiler14}, etc. In such a framework the input is not high level features, which are derived manually based on limited dataset, thus not necessarily
fitting the task at hand. Instead, the input is raw features (pixels
in image processing, characters in NLP, etc.). The DNN learns a high-level
hierarchy of the features during the training phase. The deeper the
hidden layer is the higher the abstraction level of the features (higher-level
features).

While most previous work on applying machine learning to malware analysis relied on manually crafted features, David et al. \cite{David15} trained DNN on raw dynamic analysis reports to generate malware signatures for use in a malware family classification context. In this paper we similarly train a DNN on raw dynamic analysis reports but the goal is obtaining a different functionality (APT attribution rather than signature generation).

The benefits of using raw features are: 

(1) Cheaper and less time consuming
than manual feature engineering. This is especially true in the case
of nation-state APTs, where the code requires a lot of time to reverse
engineer in-order to gain insights about features, due to obfuscation
techniques commonly used by it, as mentioned above. 

(2) Higher accuracy
of Deep Learning classifiers, since important features are never overlooked.
For instance, in our nation-state APT attribution classifier, mentioned
in the next section, we have used the technique suggested in \cite{Olden02}
to assess the contribution of each of our features, by multiplying
(and summing up) their weights in the network, where the highest value
indicates the most significant feature. We have seen that, besides
the expected API calls and IP strings of C\&C servers, arbitrary hexadecimal
values were surprisingly some of the most important features. A
security researcher might throw such addresses away, since they are
useless. However, those values were the size of data of specific PE
section which contained encrypted malicious shellcode, identifying
a specific malware family. 

(3) More flexibility due to the ability
to use the same features for different classification objectives.
For instance, our nation-state APT attribution classifier uses the
same raw features of the malware signature generator implemented in
\cite{David15}. Therefore, we could implement both, using only a
single feature extraction process. 

\section{Implementation and Experimental Evaluation}

The challenges mentioned in the previous section require a novel approach
in-order to mitigate them. As mentioned before, the problem at hand
is not a regular authorship attribution problem, since more than a
single developer is likely to be involved, some of them might be replaced
in the middle of the development. This makes regular authorship attribution
algorithms, using personal stylistic features irrelevant. Another
approach would be to consider all of the same nation state APTs as
a part of a single malware family. The rationale is that common frameworks
and functionality should exist in different APTs from the same nation.
However, this is also not accurate: each nation might have several
cyber units, each with its own targets, frameworks, functionality,
etc. Thus, it would be more accurate to look at this classification
task as a malicious/benign classifier: each label might contain several
``families'' (benign web browsers, malicious ransomware, etc.) that
might have very little in common. Fortunately, DNN is known to excel
in such complex tasks.

This brings us to the usage of raw features: since we do not know how
many ``actual APT families'' are available in the dataset, we need
to use raw features, letting the DNN build its feature abstraction
hierarchy itself, taking into account all available APT families,
as mentioned in section 1.

\subsection{Raw Features Used}

A sandbox analysis report of an executable file can provide a lot of useful
information, which can be leveraged for many different classification
tasks. In order to show the advantages of using raw features by a
DNN classifier, we have chosen raw features that can be used for different
classification tasks.

Cuckoo Sandbox is a widely used open-source project for automated
dynamic malware analysis. It provides static analysis of the analyzed
file: PE header metadata, imports, exports, sections, etc. Therefore,
it can provide useful information even in the absence of dynamic analysis,
due to, e.g., anti-VM techniques used by nation-state APTs. Cuckoo
Sandbox also provides dynamic analysis and monitors the process system
calls, their arguments and their return value. Thus, it can provide
useful information to mitigate obfuscation techniques used by nation-state
APTs. We have decided to use Cuckoo Sandbox reports as raw data for
our classifiers due to their level of detail, configurability, and
popularity. We used Cuckoo Sandbox default configuration.

Our purpose was to let our classifiers learn the high-level abstraction
hierarchy on their own, without involving any manual or domain-specific knowledge. Thus, we used words only, which are basic raw features commonly used in the
text analysis domain. Although Cuckoo reports are in JSON format, which can be parsed such that specific information is obtained from them, we did not perform any parsing. In other words, we treated the reports as raw text, completely ignoring the formatting, syntax, etc. Our goal was to let our classifiers
learn everything on their own, including JSON parsing, if necessary. Therefore,
the markup and tagged parts of the files were extracted as well. For
instance, in \textquotedbl{}api: CreateFileW\textquotedbl{} the terms
extracted are \textquotedbl{}api\textquotedbl{} and \textquotedbl{}CreateFileW\textquotedbl{},
while completely ignoring what each part means.

Specifically, our
method follows the following simple steps to convert sandbox files
into fixed size inputs to the neural network:

(1) Select as features the top 50,000 words with highest frequency in all Cuckoo reports, after removing the words which appear in all files. The rationale is that words which appear in all files, and words which are very uncommon do not contain lots of useful information.

(2) Convert each sandbox file to a 50,000-sized bit string by checking whether each of the 50,000 words appear in it. That is, for each analyzed Cuckoo report, feature{[}i{]}=1 if the i-th most common word appears
in that cuckoo report, or 0 otherwise.

In other words, we first defined which words participated in our dictionary
(analogous to the dictionaries used in NLP, which usually consist of the most frequent words in a language) and then we checked each
sample against the dictionary for the presence of each word, thus
producing a binary input vector. 

\subsection{Network Architecture and Hyper-Parameters}

We trained a classifier based on Cuckoo reports of samples of APT
which were developed (allegedly) by nation-states. Due to the small
quantity of available samples, we used only two classes: Russia and
China (which are apparently the most prolific APT developers).

Our training-set included 1,600 files from each class (training
set size of 3,200 samples) of dozens of known campaigns of nation-developed APTs. 200 samples from the training set were used as a validation set. The test set contained additional 500 files from each class (test set size of 1,000 files). The labels (i.e., ground-truth attribution) of all these files are based on well-documented and extended manual analyses within the cyber-security community, conducted during the past years.

\textbf{Note that the above-mentioned separation between training and test sets completely separates between different APT families as well. That is, if an APT family is in test set, then all its variations are also in test set only. This makes the training challenge much more difficult (and more applicable to real-world), as in many cases inevitably we will be training on APT developed by one group of developers, and testing on APT developed by a completely different group.}

Our DNN architecture is a 10-layers fully-connected neural network, with
50,000-2,000-1,000-1,000-1,000-1,000-1,000-1,000-500-2 neurons, (that is, 50,000 neurons on the input
layer, 2,000 in the first hidden layer, etc.), with an additional
output softmax layer. We used a dropout (\cite{Srivastava14}) rate
of 0.5 (ignoring 50\% of the neurons in hidden layers for each sample) 
and an input noise (zeroing) rate of 0.2 (ignoring 20\% of input neurons) to prevent overfitting. A ReLU (\cite{Glorot11}) activation function was used, and an initial learning rate of $10^{-2}$ which decayed to $10^{-5}$ over 1000 epochs. These hyper-parameters were optimized using the validation set.

\subsection{Experimental Evaluation}

Following the training phase, we tested the accuracy of the DNN model over the test set. The accuracy for the nation-state APT attribution classifer
was \textbf{94.6\%} on the test set, which contained only families that were not in the training set.

These are test accuracies are surprising in light of the complete separation of malware families between train and test sets. Inevitably in many cases the developers or even the developing units of the APT in train and test sets are different (e.g., APTs in train set developed by one Chinese cyber unit, and APTs in test set developed by another Chinese cyber unit).

Given this strict separation, and in light of the high accuracy results obtained, the results lead to the conclusion that each nation-state has (apparently) different sets of methodologies for developing APTs, such that two separate cyber units from nation A are still more similar to each other than to a cyber unit from nation B.

\section{Concluding Remarks}

In this paper we presented the first successful method for automatic APT attribution to nation-states, using raw dynamic analysis reports as input, and training a deep neural network for the attribution task. The use of raw features has the advantages of saving
costs and time involved in the manual training and analysis process. It also prevents
losing indicative data and classifier accuracy, and allows flexibility,
using the same raw features for many different classification tasks.

Our results presented here lead to the conclusion that despite all the efforts devoted by nation states and their different cyber units in developing unattributable APTs, it is still possible to reach a rather accurate attribution. Additionally, different nation-states use different APT developing methodologies, such that the works of developers in separate cyber units are still sufficiently similar to each other that allow for attribution.

While the work presented here could help facilitate automatic attribution of nation-state attacks, we are aware that nation-states could subvert the methods presented here such that they would modify their new APTs to lead to their misclassification and attribution to another nation-state. For example, using deep neural networks themselves, they could employ generative adversarial networks (GAN)\cite{Goodfellow14} to modify their APT until it successfully fools our classifier into attributing it to another nation-state. Applying GAN for APT modification would prove very difficult, but theoretically possible. 

In our future works in this area we will examine additional nation state labels (multi-class classifier), once larger datasets of such APTs become available.

\end{document}